\documentclass[draft,jgrga]{agu2001}
\usepackage{graphicx}
\usepackage{amsfonts}
\usepackage{color}
\authorrunninghead{XIONG ET AL.}
\titlerunninghead{MC--Shock Interaction and Its Geoeffectiveness 2}

\authoraddr{Ming Xiong, Huinan Zheng, Yuming Wang, and Shui Wang,
CAS Key Laboratory for Basic Plasma Physics, School of Earth and
Space Sciences, University of Science and Technology of China,
Hefei, Anhui 230026, China (mxiong@mail.ustc.edu.cn;
hue@ustc.edu.cn; ymwang@ustc.edu.cn; and swan@ustc.edu.cn)}
\begin{document}
\setkeys{Gin}{draft=false}

\title{Magnetohydrodynamic Simulation of the Interaction between Interplanetary Strong Shock and Magnetic
Cloud and its Consequent Geoeffectiveness 2: Oblique Collision}

\author{Ming Xiong, Huinan Zheng, Yuming Wang, and Shui Wang}
\affil{CAS Key Laboratory for Basic Plasma Physics, School of Earth and Space Sciences, University of
Science and Technology of China, Hefei, Anhui 230026, China}

\begin{abstract}
Numerical studies of the interplanetary ``shock overtaking
magnetic cloud (MC)" event are continued by a 2.5 dimensional
magnetohydrodynamic (MHD) model in heliospheric meridional plane.
Interplanetary direct collision (DC)/oblique collision (OC)
between an MC and a shock results from their same/different
initial propagation orientations. For radially erupted MC and
shock in solar corona, the orientations are only determined
respectively by their heliographic locations. OC is investigated
in contrast with the results in DC \citep{Xiong2006}. The shock
front behaves as a smooth arc. The cannibalized part of MC is
highly compressed by the shock front along its normal. As the
shock propagates gradually into the preceding MC body, the most
violent interaction is transferred sideways with an accompanying
significant narrowing of the MC's angular width. The opposite
deflections of MC body and shock aphelion in OC occur
simultaneously through the process of the shock penetrating the
MC. After the shock's passage, the MC is restored to its oblate
morphology. With the decrease of MC-shock commencement interval,
the shock front at 1 AU traverses MC body and is responsible for
the same change trend of the latitude of the greatest
geoeffectiveness of MC-shock compound. Regardless of shock
orientation, shock penetration location regarding the maximum
geoeffectiveness is right at MC core on the condition of very
strong shock intensity. An appropriate angular difference between
the initial eruption of an MC and an overtaking shock leads to the
maximum deflection of the MC body. The larger the shock intensity
is, the greater is the deflection angle. The interaction of MCs
with other disturbances could be a cause of deflected propagation
of interplanetary coronal mass ejection (ICME).
\end{abstract}
\begin{article}


\section{Introduction}\label{Sec:Intro}
Interplanetary (IP) space is permeated by highly fluctuating solar
wind with magnetic field frozen in its plasma \citep{Parker1963}.
The relatively quiet equilibrium of IP space is frequently
interrupted by the solar disturbances, especially during solar
maximum. Giant clouds of ionized gas with magnetic flux of
$10^{23}$ maxwell and plasma mass of $10^{16}$ g, called coronal
mass ejection (CME), are regularly emitted from the sun
\citep{Gosling1990,Webb1994}. IP CME (ICME) generally causes
strong perturbation in the space environment as it passes by.
Several models have already been applied in space weather
forecasting, such as (1) HAF (Hakamada-Akasofu-Fry)
\citep{Fry2001,Fry2005}; (2) STOA (Shock Time of Arrival)
\citep{Smart1985}; (3) ISPM (Interplanetary Shock Propagation
Model) \citep{Smith1990}; (4) an ensemble of HAF, STOA and ISPM
models \citep{Dryer2001,Dryer2004}; (5) SWMF (Space Weather
Modeling Framework) \citep{Groth2000,Gombosi2001,Toth2005}; (6)
HHMS (Hybrid Heliospheric Modeling System) \citep{Detman2006} ,
and so on. Great challenges are still faced to improve the
prediction performance of space weather to satisfy the
ever-increasing demands from human civilization \citep{Baker2002}.

Magnetic clouds (MCs) are an important subset of ICMEs, whose
fraction decreases from $\sim$ 100\% (though with low statistics)
at solar minimum to $\sim$ 15\% at solar maximum
\citep{Richardson2004,Richardson2005}. Identified by their
characteristics including enhanced magnetic field, large and
smooth rotation of magnetic field and low proton temperature
\citep{Burlaga1981}, MCs have been the subject of increasingly
intense study. The MCs with long interval of large southward
magnetic field $B_s$ are widely considered to be the major IP
origin of moderate to intense geomagnetic storms, especially
during the solar maximum
\citep{Tsurutani1988,Gosling1991,Gonzalez1999} and, hence, play a
crucial role in space weather prediction. An MC should probably be
a curved loop-like structure with its feet connecting to the solar
surface \citep{Larson1997}. The force-free magnetic flux rope
models have been proven to be very valuable to interpret in situ
observations of MCs
\citep{Lundquist1950,Goldstein1983,Burlaga1988,Farrugia1993}. For
the study of evolution of an individual MC during its anti-sunward
propagation, many sophisticated models are developed based on
these initial flux rope models: (1) Analytical models
\citep{Osherovich1993a,Osherovich1993b,Osherovich1995,Hidalgo2003,Hidalgo2005};
(2) Kinematic models \citep{Riley2004,Owens2006}; (3) Numerical
models \citep{Vandas1995,Vandas1996,Vandas1997,
Vandas2002,Groth2000,Odstrcil2002,Schmidt2003,Manchester2004a,Manchester2004b}.
Especially numerical simulations in (3) on a single MC have been
exhaustive under the condition of various magnetic field
strengths, axis orientations and speeds.

ICME is not an absolutely self-isolated entity during IP
propagation. It may interact with other solar transients (e.g.,
shock, ejecta) and heterogenous medium (e.g., corotating
interacting region). With less defined characteristics, some IP
complex structures are reported recently, such as complex ejecta
\citep{Burlaga2002}, multiple MCs \citep{Wang2002a,Wang2003a},
shock-penetrated MC \citep{Wang2003b,Berdichevsky2005},
non-pressure-balanced ``MC boundary layer'' associated with
magnetic reconnection \citep{Wei2003,Wei2006}, ICME compressed by
the following high-speed stream \citep{Dal Lago2006}, and so on.
Dynamical response and ensuing geoeffectiveness of these
structures are directly associated with the interaction during
their formation and evolution. Numerical simulations have been
applied to study most of the complex structures: e.g., the
interaction of a shock wave with an MC
\citep{Vandas1997,Odstrcil2003,Xiong2006}, and the interaction of
two MCs
\citep{Odstrcil2003,Gonzalez-Esparza2004,Lugaz2005,Wang2005}.

The observed ``shock overtaking MC" events substantiate the
likelihood of strong shock propagation in low $\beta$ medium of MC
plasma and, therefore, present a very interesting topic in IP
dynamics. The evolution stages of MC-shock interaction within 1 AU
are determined by MC and shock commencement interval in solar
corona. They can be assorted into two categories: (1) shock still
in MC (e.g. October 3-6 2000 and November 5-7 2001 events
\citep{Wang2003b}); (2) shock ahead of MC after completely
penetrating it (e.g. March 20-21 2003 event
\citep{Berdichevsky2005}). The idea that shock compression of the
preexisting southward magnetic component can increase
geoeffectiveness of the corresponding $B_s$ event has been proved
in data analyses \citep{Wang2003d}. Particularly, MC-shock
compounds in  category (1) cause highly intense geomagnetic storms
\citep{Wang2003b,Wang2003c,Xiong2006}. Furthermore the
geoeffectiveness variance of MC-shock compound with respect to the
increasing depth of a shock entering a preceding MC was
investigated in our previous study \citep[][hereinafter referred
to as paper 1]{Xiong2006}. Both MC core and shock nose are
radially erupted along heliospheric current sheet (HCS) in paper
1; however, the above-mentioned specific MC-shock events
\citep{Wang2003b,Berdichevsky2005} were all identified such that
the shock flank sweeps the preceding MC body. IP direct collision
(DC)/oblique collision (OC) of an MC and a shock results from
their same/different initial propagation orientation. For radially
erupted MC and shock in solar corona, the orientations are only
determined respectively by the heliographic locations of MC core
and shock nose. Because the probability of MC core and shock nose
radially launching from the same heliographic location is very
rare and shock front extends over a wide angular span in IP
medium, it is meaningful to study the role of shock orientation
relative to a preceding MC propagation. DC in paper 1 is here
modified to be OC for MC-shock interaction. The shock in DC/OC is
correspondingly named as ``central''/``non-central'' shock.
Moreover DC/OC is likely to be the IP interaction of two radially
propagating disturbances from the same/different solar activity
regions.

Section \ref{Sec:Method} presents a brief description of numerical magnetohydrodynamic (MHD) model. Section
\ref{Sec:MCShock} discusses the dynamical evolution of MC-shock OC.  Section \ref{Sec:Geoeffect} analyzes
the ensuing geoeffectiveness of MC-shock compound. Section \ref{Sec:Deflect} describes the dependence of
shock-induced MC deflection on shock orientation and intensity. Section \ref{Sec:Conclusion} summaries the
conclusions.

%
%
\section{Numerical MHD Model}\label{Sec:Method}
The detailed description of the numerical model, including
numerical scheme, computational mesh layout, prescription of the
ambient solar wind and preceding MC, is given in paper 1. Only the
shock introduction among input parameters of numerical model is
modified to simulate OC of MC-shock interaction in contrast with
DC in paper 1.

An incidental fast shock, which is radially launched from the
inner boundary, is prescribed by several parameters: its emergence
time $t_{s0}$, the latitude of its nose $\theta_{sc}$, the
latitudinal width of its flank $\Delta \theta_s$, the maximum
shock speed within its front $v_s$, the duration of growth,
maintenance and recovery phases ($t_{s1}$, $t_{s2}$, $t_{s3}$).
Some parameters are fixed in all simulation cases of paper 1 and
here, i.e. $\Delta\theta_s = 6^\circ, t_{s1} = 0.3 \mbox{ hour},
t_{s2} = 1 \mbox{ hour}, t_{s3} = 0.3 \mbox{ hour}$. The remaining
parameters ($t_{s0}$, $\theta_{sc}$, $v_s$) are independently
chosen to mimic different conditions of IP MC-shock interaction.
$t_{s0}$ is used to separate the MC and shock initialization in
time for reproducing different evolutionary stages of MC-shock
compound at 1 AU. $\theta_{sc}$ designates emergence orientation
of shock nose relative to previous MC propagation. Since the
preceding MC emerges from the heliospheric equator,
$\theta_{sc}=0^\circ$ and $\theta_{sc} \neq 0^\circ$,
corresponding to the introduction of ``central" and ``non-central"
shock, determine MC-shock DC and OC in IP space respectively.
$v_s$ describes the intensity of MC-shock interaction to some
extent. All introduced shocks in our simulation are strong enough
to be faster than the local magnetosonic speed at all time and,
therefore, to prevent weak shock dissipation in MC medium.

\section{Dynamics of MC-shock Interaction}\label{Sec:MCShock}
All fifty simulation cases are assorted into five groups in
Table~\ref{Tab1}. Groups of individual MC (IM), direct collision
(DC), oblique collision (OC), shock orientation dependence (SOD),
and shock intensity dependence (SID) are studied respectively,
where Groups IM and DC have been addressed in detail in paper 1.
Case P$_1$ is shared by Groups DC and SOD, and Case P$_2$ shared
by Groups OC, SOD and SID. With the identical $v_s$ of $1630
\mbox{ kms}^{-1}$ and variable $t_{s0}$ from $3 \mbox{ hours}$ to
$41 \mbox{ hour}$, Groups DC and OC only differ in $\theta_{sc}$
for comparative study. By modifying $\theta_{sc}$ from $0^\circ$
to $10^\circ$, ``central" shock in DC is directed to be
``non-central" one in OC. Further, the parametric studies of
$\theta_{sc}$ from $0^\circ$ to $45^\circ$ in Group SOD and $v_s$
from $947 \mbox{ kms}^{-1}$ to $3173 \mbox{ kms}^{-1}$ in Group
SID are explored as a supplement to Groups DC and OC. Cases B$_1$
and B$_2$ with $t_{s0}=41$ hours, and Cases C$_1$ and C$_2$ with
$t_{s0}=10$ hours are typical examples of MC-shock interaction in
categories 1 and 2 referred in Section \ref{Sec:Intro}.

\subsection{Case B$_2$}
\begin{figure}
   \includegraphics[height=39pc,angle=90]{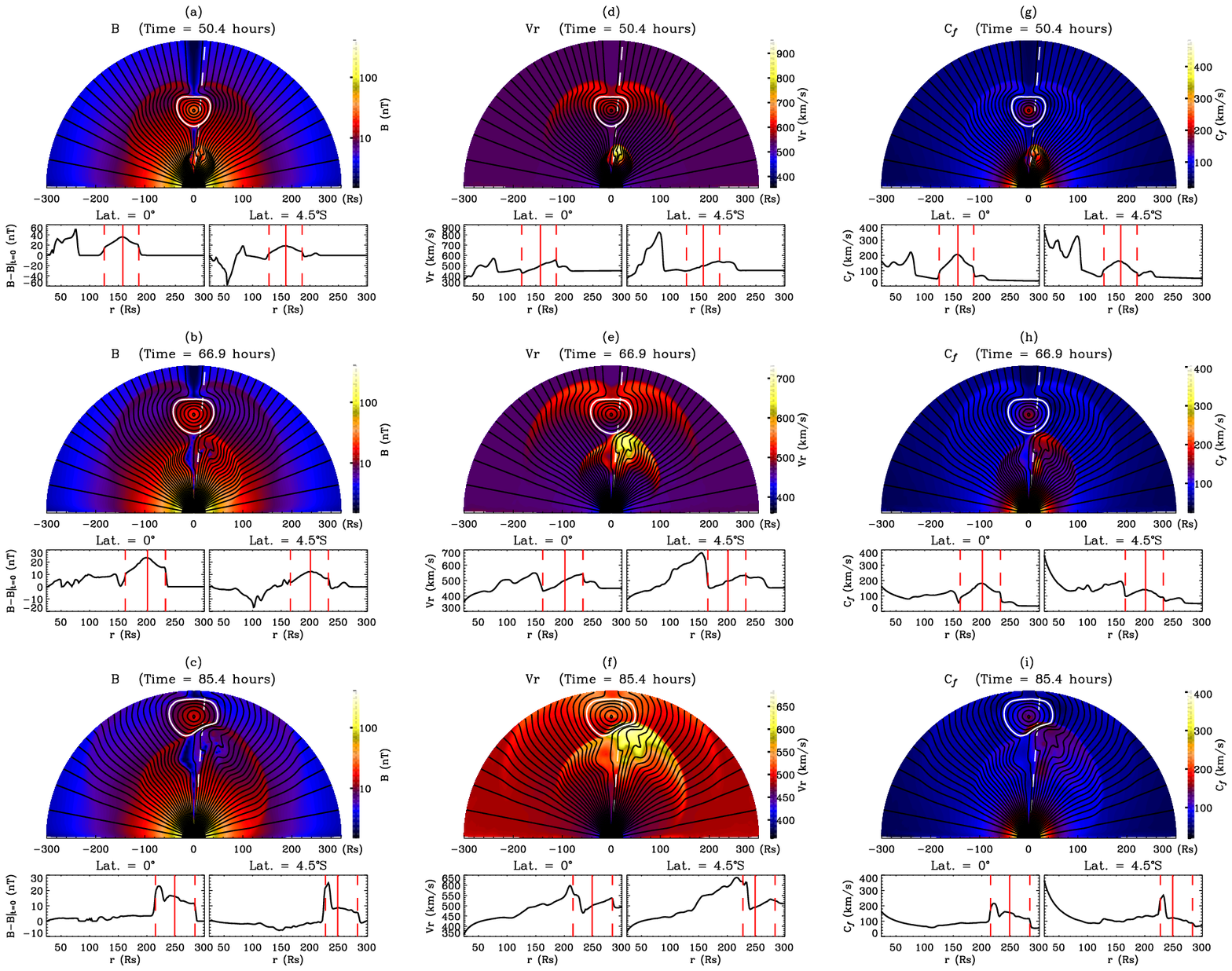}
\caption{} \label{case-B}
\end{figure}

The process of MC-shock interaction of Case B$_2$ is visualized in
Figure~\ref{case-B}. Under each image are two corresponding radial
profiles by cutting right through $0^\circ$ (noted by
$~\mbox{Lat.}=0^\circ$) and southern $4.5^\circ$ (white dashed
lines in the images, noted by $~\mbox{Lat.}=4.5^\circ$S) away from
the equator. The magnitude of magnetic field in radial profile is
given by subtracting its corresponding initial value of ambient
equilibrium. The body of MC is identified to be enclosed by a
white solid line in the images and between two dotted lines in
attached profiles. Magnetic field configuration is superimposed
upon the images. The incidental shock aphelion arrives at $90R_s$
(along Lat. $=4.5^\circ$S) in 50.4 hours meanwhile the MC core
arrives at $160R_s$ (along Lat. $=0^\circ$), shown in
Figure~\ref{case-B}(a), (d) and (g). Impending collision can be
pregnant from large radial speed difference between the preceding
MC and the following shock, as indicated by radial bulk flow speed
$v_r$ of $830 \mbox{ kms}^{-1}$ at shock front and $540 \mbox{
kms}^{-1}$ at MC head from the profile of Lat. $=4.5^\circ$S
(Figure~\ref{case-B}(d)). Though the latitudinal span of its flank
is $6^\circ$ initially at inner boundary, the shock extends up to
$40^\circ$ quickly due to its very strong intensity, until it
emerges into IP medium completely. The traverse of shock front
across the equator leads to significant HCS warping seen clearly
in Figure~\ref{case-B}(b), which is consistent with previous
results \citep{Smith1998,Hu2001}. As shock emergence orientation
is redirected, the morphology of IP shock changes from a concave
(Figure 3(e) in paper 1) to a smooth arc (Figure~\ref{case-B}(e)
here). As a result, MC-shock interaction consequently changes from
DC to OC. The shock just catches up with the inner boundary of MC
at 66.9 hours (Figure~\ref{case-B}(b), (e) and (h)). Due to strong
magnetic field and low $\beta$ plasma, the radial characteristic
speed of fast mode wave $c_f$ of the MC is abnormally high at 1 AU
with $200 \mbox{ kms}^{-1}$ in maximum at MC core and $100 \mbox{
kms}^{-1}$ in minimum at MC boundary. The rare chance of shock
survival in an MC medium explains why only a few ``shock
overtaking MC" events are observed in IP space. Across the tangent
point between inner MC boundary and shock front exists a quite
sharp slope of $v_r$, as clearly seen along Lat. $=4.5^\circ$S.
MC-shock interaction begins from this tangent point at 66.9 hours.
Once a slow MC is within the very large latitudinal span of the
overtaking shock front, it will be swept by the shock and, from
then on, the evolution of MC and shock will be coupled with each
other. The overwhelming shock significantly distorts MC morphology
at 85.4 hours (Figure~\ref{case-B}(c), (f) and (i)). Namely, the
originally curved magnetic field lines become very flat. The
collision is more violent along Lat. $=4.5^\circ$S. A sharp
discontinuity is conspicuously formed in the rear part of MC with
$B-B|_{t=0}=25 \mbox{ nT}$, $v_r=620 \mbox{ kms}^{-1}$, and
$c_f=260 \mbox{ kms}^{-1}$ in maximum within highly compressed
region.

\subsection{Case C$_2$}\label{Sec:CaseC}
\begin{figure}
\noindent
    \includegraphics[width=20pc]{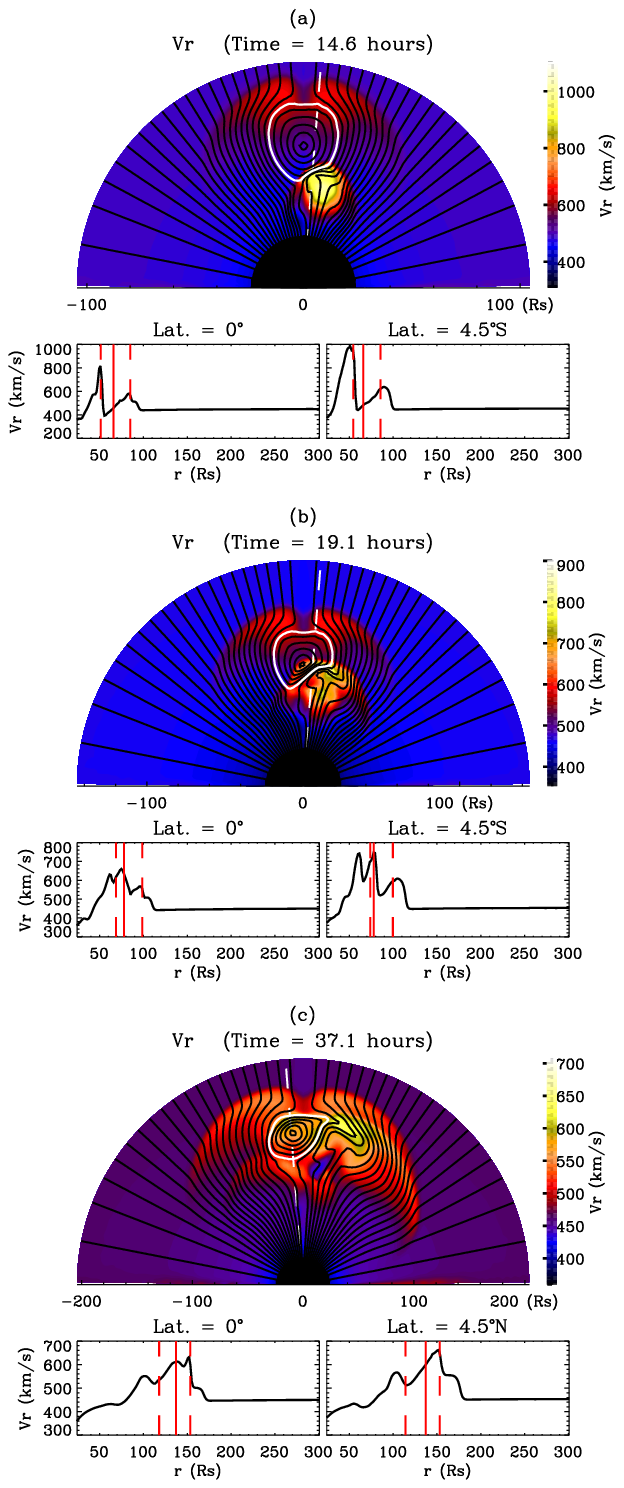}
\caption{} \label{Case-C}
\end{figure}

In Case C$_2$, an earlier shock emergence ($t_{s0}=10$ hours)
allows the incidental shock to ultimately penetrate the MC body
within the solar-terrestrial heliospheric range. Only the
evolution of $v_r$ is given in Figure~\ref{Case-C} to show the
concerned MC-shock complex structure. Though an MC generally
behaves like a rigid body with a little elasticity, magnetic field
lines of the simulated MC appear to be too vulnerable to be easily
deformed in the face of an overwhelming shock. The shock is
radially emitted with the strongest intensity at front nose. Hence
shock front behaves as an oblique curve relative to heliospheric
equator due to the propagation speed difference from shock nose to
edge flank. The MC is highly compressed by the shock along its
normal. The shock front looks like a smooth arc in MC medium. As
it propagates gradually into the preceding MC body, the most
violent interaction is transferred sideways (heliolatitudinally in
the present study). Due to net shock-input angular momentum during
MC-shock OC, the MC core starts to deflect away from initial shock
orientation when the shock enters MC core, as seen in the contrast
of Figure~\ref{Case-C}(b) and (c). The overall MC body is also
deflected to the north. The global MC body deflection is
quantified by the deflection angle of its core. Once the shock
completely penetrates the MC, the grip of shock force on the MC is
substantially relaxed, and the MC is restored to the roughly
ellipse morphology by its field line elasticity. Meanwhile, the MC
loses its angular speed component by the relative difference
between the radial ambient flow and the speed's value at the MC
boundary and, then propagates radially along the deflected angle.
The incidental shock is also simultaneously deviated with its
aphelion in the opposite direction, until it finally merges with
the MC-driven shock into a compound one. The bend of
interplanetary magnetic field (IMF) lines is obvious near the
south of MC boundary, seen from Figure~\ref{Case-C}(c).

\begin{figure}
   \includegraphics[scale=0.7]{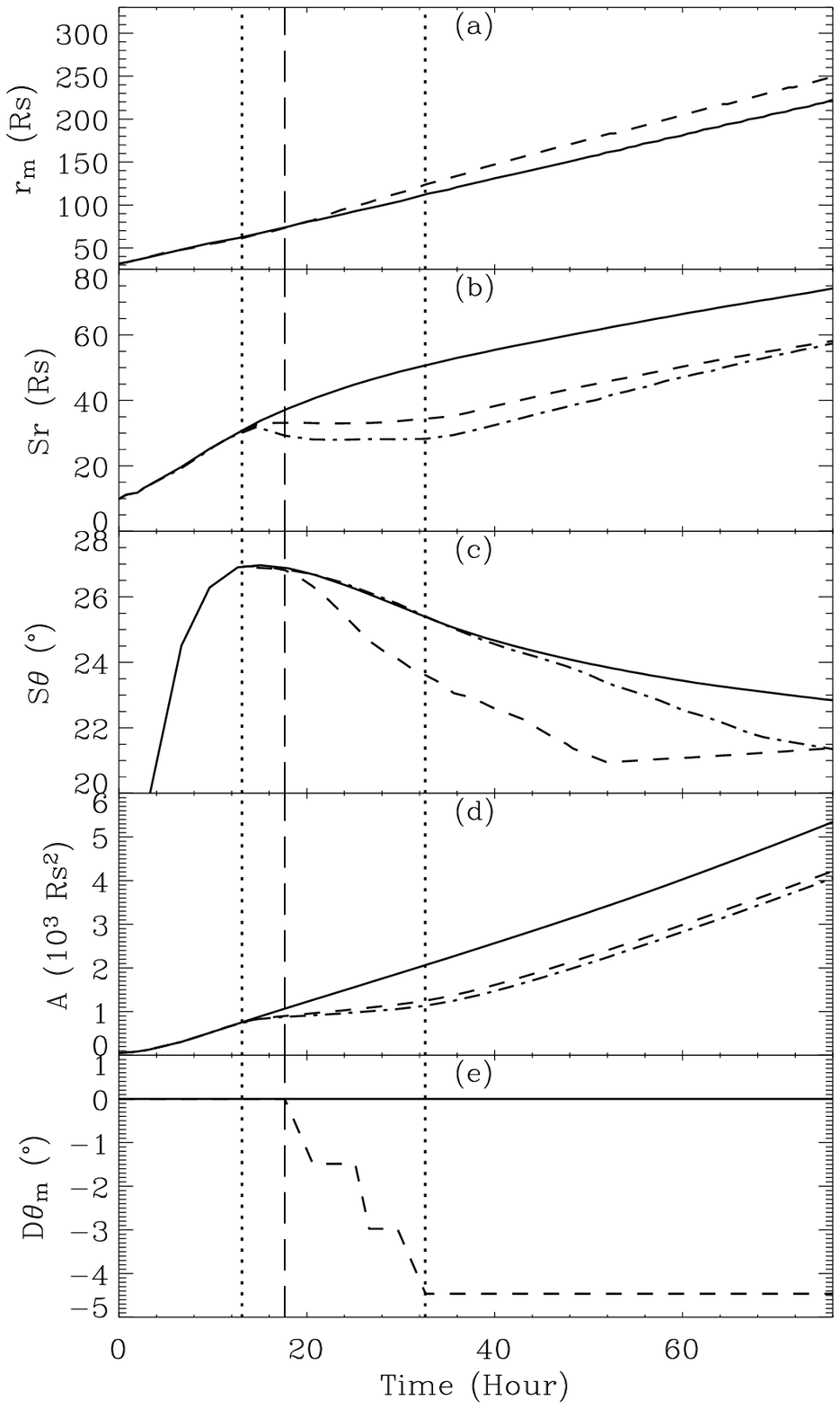}
\caption{} \label{multi-geometry}
\end{figure}

Figure~\ref{multi-geometry} shows the comparison among Cases A, C$_1$, and C$_2$ about time-dependent
parameters: (a) radial distance of MC core $r_m$, (b) MC radial span $Sr$, (c) MC angular span $S\theta$,
(d) MC cross section area $A$, and (e) MC core deflection angle $D\theta_m$, where the solid, dashed and
dashed-dotted curves denote Cases A, C$_2$ and C$_1$ respectively, and the three vertical delimiting lines
(dotted, dashed and dotted) from left to right correspond to the occasion of shock encountering MC tail,
core and head respectively. The MC in Case C$_2$ is largely compressed by the shock, beginning from 13
hours. The dependence of the compression of MC geometry on shock orientation is illustrated by the
comparison in Figure~\ref{multi-geometry}(b)-(d). $Sr$ is larger while $S\theta$ is smaller for Case C$_2$
in Group OC. Though $S\theta$ is little affected in Case C$_1$ when shock front is in MC body ($13$ hrs $ <
t < 33$ hrs), it is significantly narrowed in Case C$_2$. And the MC cross section area $A$ in Case C$_2$,
which represents the overall influence of shock compression due to integration of factor $Sr$ and
$S\theta$, is a bit larger than that in Case C$_1$. Starting from being encountered by the following shock,
MC core deflects up to $-4.5^\circ$ until shock front reaches MC head, as seen in
Figure~\ref{multi-geometry}(e). Though total deflection angle of MC core ($-4.5^\circ$) amounts to three
computational grids of latitudinal spacing $1.5^\circ$, MC deflection, we think, is indeed physical
solution. Due to rough subcell resolution in numerical computation, MC core deflection behaves as a false
discrete quantum-like transition instead of a realistic smooth one. But it does not distort the fundamental
physical characteristics in numerical simulation.

\subsection{Multi-Cases Comparison}
\begin{figure}
   \includegraphics[width=20pc]{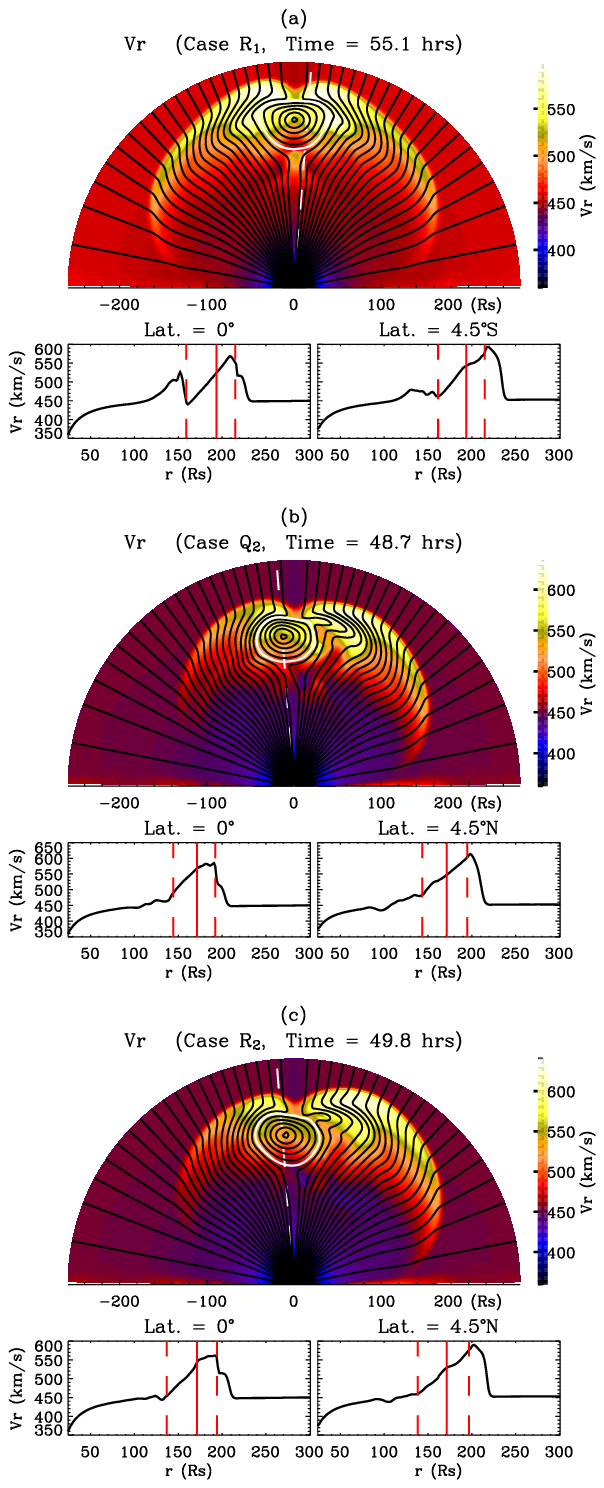}
\caption{} \label{montage}
\end{figure}

The propagation of MC-shock structure toward the earth can be
detected by L1-orbiting spacecraft, which perform the sentinel
duty in space weather alarm system. The montage of the evolution
of MC-shock compound at L1 under three typical circumstances is
visualized in Figure~\ref{montage}, where (a)-(c) correspond to
Case R$_1$ from Group DC and Cases Q$_2$ and R$_2$ from Group OC.
Though the farthest radial distances of shock front in the north
and south of the equator are almost identical in Cases R$_1$ and
Q$_2$, the shock intensity in the south in Case Q$_2$ is
apparently stronger than its north counterpart. With a smaller
emergence interval, the shock in Case R$_2$ merges completely with
the MC-driven shock into a compound one and moves faster in the
south by contrast of Figure~\ref{montage}(b) and (c). Moreover the
asymmetry of compound shock front with respect to heliospheric
equator occurs when the shock erupts sideways relative to the MC
propagation. The final MC propagation is slightly deviated from
heliospheric equator to northern $4.5^\circ$ after being
ultimately penetrated by the shock, as seen from
Figure~\ref{montage}(b) and (c). The succedent high speed flow
right after the inner boundary of preceding MC in Group DC,
mentioned in paper 1, does not exist in corresponding Group OC,
which can be seen from contrast between Figure~\ref{montage}(a),
(b), and (c). The shock front with $\theta_{sc} \neq 0^\circ$ has
the oblique normal relative to the preceding MC propagation, so
the disturbance of speed enhancement downstream of shock front in
Group OC can completely bypass or penetrate the obstacle of MC
body and merge with the MC-driven shock.

\section{Geoeffectiveness Studies}\label{Sec:Geoeffect}
The southward magnetic flux within the MC is located in its rear part. The geomagnetic effect of simulated
$B_s$ event is quantified by $Dst$ index. The in-situ measurements by a hypothetic spacecraft at L1 are
inputted to Burton formula \citep{Burton1975} to calculate $Dst$, as applied by
\citet{Wang2003c,Xiong2006}.

\begin{figure}
   \includegraphics[width=20pc]{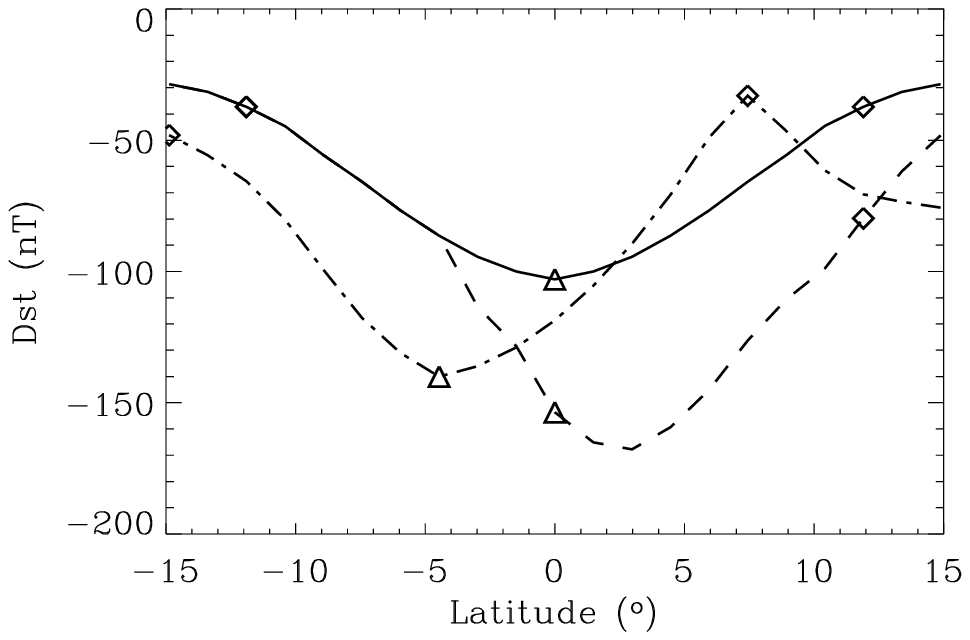}
\caption{} \label{multi-lat}
\end{figure}

Near-HCS latitudinal dependence of $Dst$ index in Cases A, B$_2$, and C$_2$ is plotted in
Figure~\ref{multi-lat}. The positive and negative latitudes are referred to southern and northern
semi-heliosphere. With the MC core marked by $\Delta$ and MC boundary by $\diamond$, the solid, dashed, and
dashed-dotted lines denote Cases A, B$_2$, and C$_2$ respectively. Geomagnetic storm has been obviously
aggravated by shock overtaking MC. The minimum $Dst$ are found to be -103 nT, -168 nT, and -140 nT in Cases
A, B$_2$, and C$_2$, respectively. Cases B$_2$ and C$_2$ are discussed one by one against Case A. Firstly,
geomagnetic storm in Case A is largely enhanced in Case B$_2$ within the latitudinal span influenced by the
shock. The minimum $Dst$ occurs at $3^\circ$ rather than $0^\circ$ (the latitude of MC core passage),
because the former undergoes more violent compression. The geoeffectiveness remain unchanged within
$\mbox{Lat.} < -5^\circ$. The asymmetry of shock propagation with respect to heliospheric equator leads to
subsequent asymmetry of geoeffectiveness of the MC-shock compound. Secondly, in Case C$_2$ the concave of
the latitudinal distribution of $Dst$ is shifted $4.5^\circ$ to the north. The MC deflection is caused by
``non-central" shock penetrating MC body, as interpreted in section~\ref{Sec:CaseC}. As a result, the southward passing
magnetic flux decreases due to the northward deflection of MC, and the IMF bend south of the equator due
to shock passage, seen from Figure~\ref{Case-C}(c), which are responsible for the increased and decreased
$Dst$ in $2.3^\circ < \mbox{Lat.} < 9.4^\circ$ and $9.4^\circ < \mbox{Lat.} < 15^\circ$ respectively,
comparing with Case A. Therefore, as shock front propagates from the south (Case B$_2$) to the north (Case
C$_2$) in MC medium, the latitude of minimum $Dst$ consequently moves in the same direction.

\begin{figure}
   \includegraphics[width=20pc]{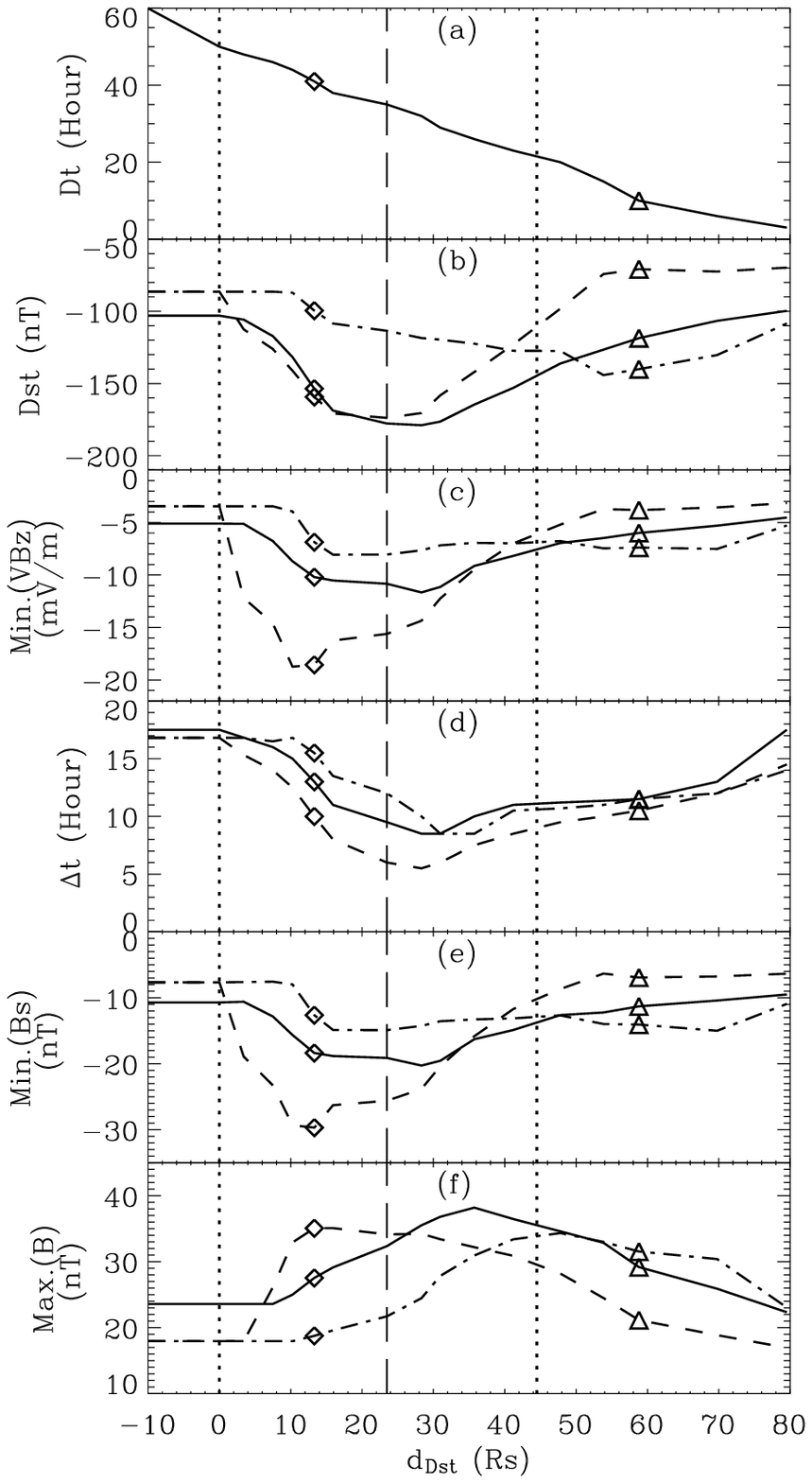}
\caption{} \label{depth}
\end{figure}

All MC-shock interaction cases of Group OC are integrated to study further the dependence of $Dst$ index on
the penetration depth $d_{Dst}$ of shock overtaking MC. $d_{Dst}$ is defined as the radial distance between
shock front and MC inner boundary along sun-MC core. Three in-situ observations in time sequence at L1
along heliospheric equator and $\pm 4.5^\circ$ aside are synthetically analyzed in Figure~\ref{depth},
where the three vertical delimiting lines (dotted, dashed and dotted) from left to right correspond to the
cases of shock encountering the tail, the core and the front of MC at L1, respectively. From top to bottom
are plotted (a) MC-shock emergence interval, noted by $Dt$, (b) the $Dst$ index, (c) the minimum of
dawn-dusk electric field $VB_z$, noted by $~\mbox{Min.}(VB_z)$, (d) the interval between the commencement
of $VB_z < -0.5~\mbox{ mV/m}$ and the corresponding $Dst$ minimum, noted by $\Delta t$, (e) the minimum of
southward magnetic component $B_s$, noted by $\mbox{Min.}(B_s)$, and (f) the maximum of magnetic field
magnitude $\mbox{Max.}(B)$, respectively. The solid, dashed and dashed-dotted lines in
Figure~\ref{depth}(b)-(f) correspond to the observations at $\mbox{Lat.=}0^\circ$, $4.5^\circ$S and
$4.5^\circ$N, respectively. The separate MC and shock events are coupled together when $Dt<$ 50 hours. The
shock penetrates into the preceding MC more deeply with shorter $Dt$. $\mbox{Min.}(B_s)$ and
$\mbox{Min.}(VB_z)$ decline dramatically along Lat. = $4.5^\circ$S as $d_{Dst}$ increases from 0 to
$10R_s$, because the first tangent point between MC boundary and shock front is very near $4.5^\circ$S.
$Dst$ decreases monotonically within $0 R_s < d_{Dst} < 23.5 R_s$ until shock front reaches MC core. Once
the shock front exceeds the MC core ($d_{Dst} > 23.5 R_s$), the latter begins to deflect northward.
Moreover when $d_{Dst} > 23.5 R_s$, the greatest compression region by the shock front is within the MC
anterior part or the MC-driven sheath, where magnetic field is northward and, hence, contributes little to
geoeffectiveness. So the mitigated geoeffectiveness along $0^\circ$, $4.5^\circ$S and aggravated
geoeffectiveness along $4.5^\circ$N coexist, as seen from $23.5 R_s < d_{Dst} < 44.5 R_s$ in
Figure~\ref{depth}(b).

\begin{figure}
   \includegraphics[width=20pc]{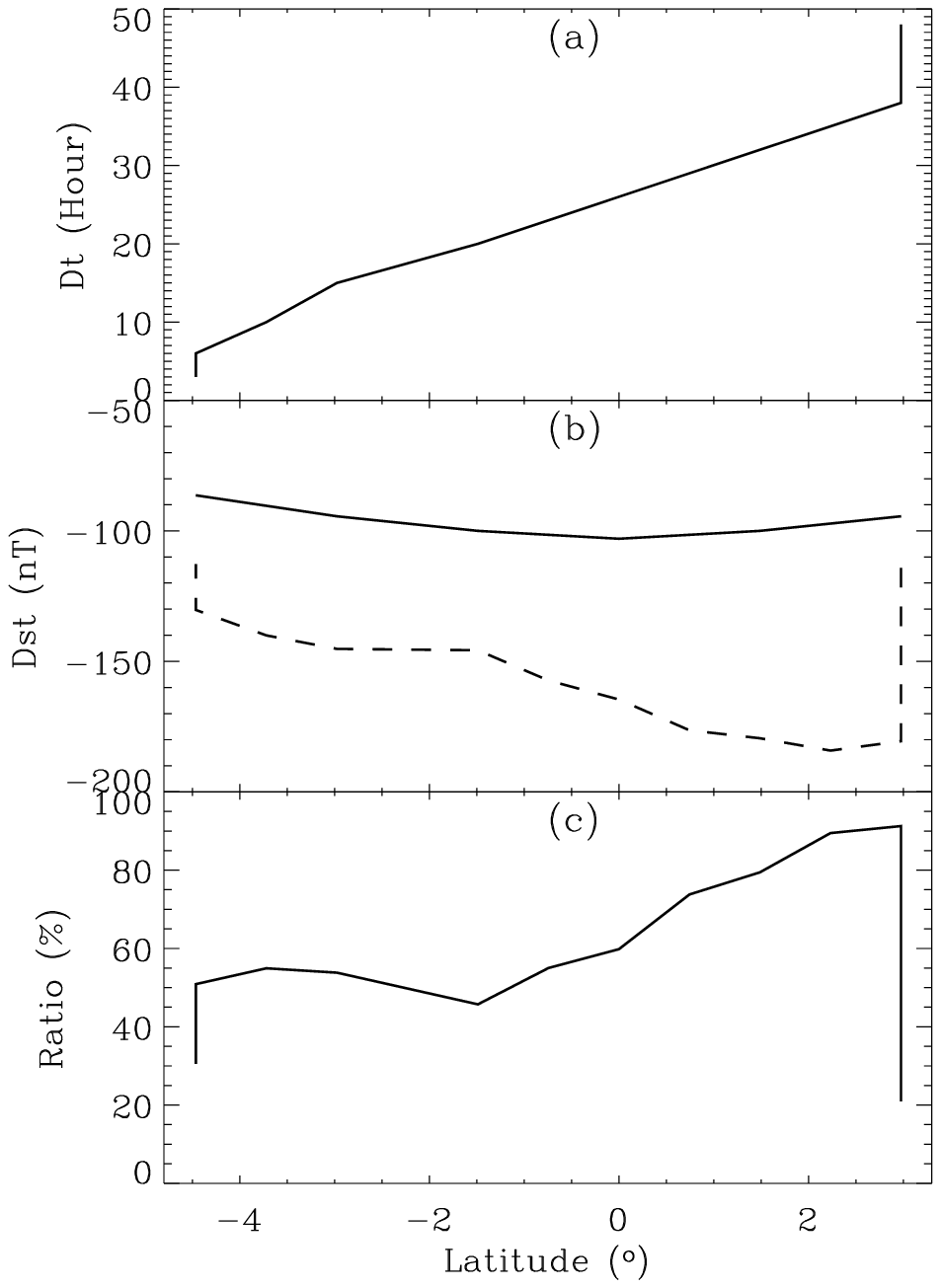}
\caption{} \label{multi-minLat}
\end{figure}

Based on the analyses of Figures~\ref{multi-lat} and \ref{depth},
MC deflection by MC-shock OC plays a crucial role in geomagnetic
storms. The minimum $Dst$ and its corresponding latitude among
$Dst$ latitudinal distribution for every case of Group OC are
assembled in Figure~\ref{multi-minLat}. With a given $Dt$, there
exists a latitude where geoeffectiveness reaches its maximum
(Figure~\ref{multi-minLat} (a)). This specific $Dst$ value is
plotted as dashed line in Figure~\ref{multi-minLat}(b). The
latitudinal distribution of individual MC event (Case A), serving
as a background in contrast, is also plotted as solid line in
Figure~\ref{multi-minLat}(b). The relative ratio of
geoeffectiveness enhancement by the shock is presented in
Figure~\ref{multi-minLat}(c) to quantify two curves difference in
Figure~\ref{multi-minLat}(b). As $Dt$ decreases from 48 hours to 3
hours, the latitude of maximum geoeffectiveness firstly remains
constant with decreased $Dst$ from -115 nT to -180 nT, enhanced
ratio from 20\% to 91\%, then monotonically changes from $3^\circ$
to $-4.5^\circ$ with gradually subdued geoeffectiveness, finally
remains constant again with further increased $Dst$ from -130 nT
to -115 nT, decreased ratio from 50\% to 30\%. The minimum $Dst$
(-185 nT) occurs at $2.3^\circ$ when the shock front enters MC
core right at 1 AU. In contrast with paper 1, the maximum
geoeffectiveness of MC-shock interaction in Group DC is the same
as that in Group OC despite occurrence at different
heliolatitudes.

\section{MC and Shock Deflections}\label{Sec:Deflect}
\begin{figure}
   \includegraphics[width=39pc]{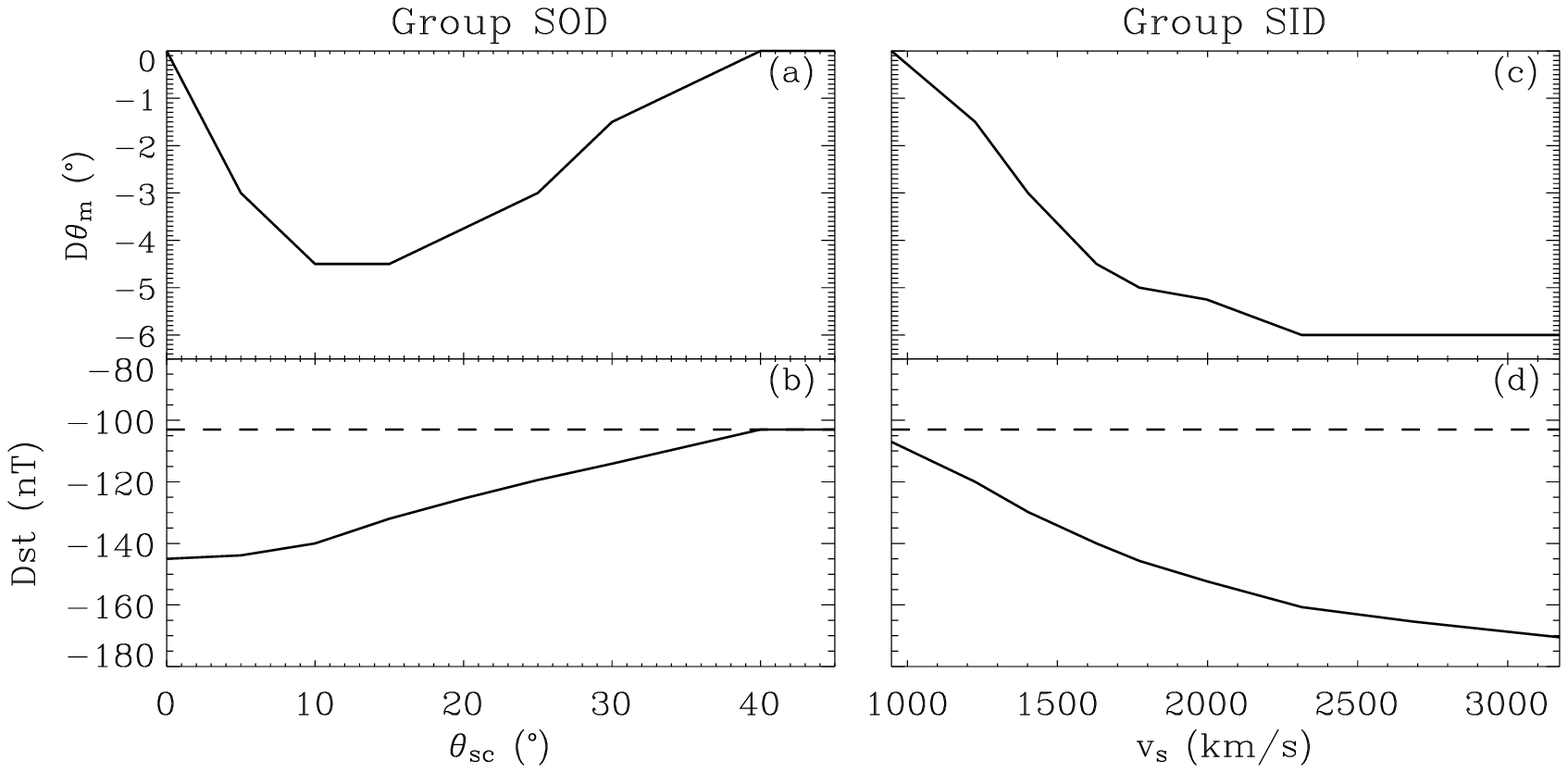}
\caption{} \label{SOD-SID}
\end{figure}

IP MC deflection mentioned in Section~\ref{Sec:CaseC} is a key parameter for solar-terrestrial
transportation process, because it concerns the preexisting condition of geomagnetic storms -- whether an
MC could encounter the earth. In order to explore reliance of MC core deflection angle on shock orientation
and intensity, the results of Groups SOD and SID are illustrated in Figure~\ref{SOD-SID}. Because MC core
continuously deflects on the condition of shock front being in MC medium, seen from
Figure~\ref{multi-geometry}(e), all $t_{s0}$ in Groups SOD and SID are chosen to be 10 hours to have MC
completely penetrated for obtaining final invariant angular displacement of MC core $D\theta_m$. $Dst$ in
Figure~\ref{SOD-SID} refers to the geoeffectiveness at certain latitude of passage of deflected MC core.
Firstly for Group SOD with different $\theta_{sc}$, two factors affect $D\theta_m$: (1) $\theta_{sc} \neq
0$ is a premise of MC core deflection; $D\theta_m = 0$ corresponds to $\theta_{sc} = 0$. (2) As
$\theta_{sc}$ increases, shock flank section encountered by MC body is further away from shock nose and,
hence weaker. The absolute value of deflection angle tends to be smaller due to the weakening of MC-shock
collision. The maximum deflection of MC core ($D\theta_m=-4.5^\circ$) occurs at certain $\theta_{sc}$
($10^\circ < \theta_{sc} < 15^\circ$). Meanwhile $Dst$ increases monotonically as a function of
$\theta_{sc}$, up to the value of corresponding individual MC event. Secondly for Group SID with different
$v_s$, both $D\theta_m$ and $Dst$ decrease steadily as $v_s$ increases. Moreover, the slopes of two curves
in Figure~\ref{SOD-SID}(c) and (d) decrease steadily, very abrupt in $v_s=1000$ km/s and become nearly
horizontal when $v_s \ge 3000$ km/s. This saturation effect on $D\theta_m$ and $Dst$ is caused by the
concurring deflection of shock aphelion opposite to that of MC core mentioned in Section~\ref{Sec:CaseC}.
So the divergent trend of deflection angle between the MC body and the shock aphelion counteracts, more or
less, the effect of increasing shock speed $v_s$ on MC-shock collision.

The finding of MC deflection due to interaction with a shock is
further discussed through comparison with other relevant models.
(1) \citet{Vandas1996} proposed that an MC deflects during the
propagation through IP medium with unipolar IMF. Magnetic
reconnection between IMF and inherent MC field across one side of
MC boundary causes the angular force unbalance and, hence, leads
to angular deflection. The MC continuously deflects through IP
space. The role of magnetic helicity is responsible for deflection
mechanism \citep{Vandas1996}. However, such deflection needs to be
verified further, as the reconnection should not be so significant
in the IP medium with low $\beta$; (2) \citet{Wang2004} suggested
that CMEs could be deflected as largely as several tens degrees in
the propagation under the effects of background solar wind and
spiral IMF. CME deflects from its onset until accelerated or
decelerated to background solar wind, which is expected to be done
within several tens solar radii \citep{Wang2006b}. It can well
interpret the observation fact of east-west asymmetry of solar
source distribution of earth-encountered halo CMEs
\citep{Wang2002b} and why some eastern limb CMEs encountered the
earth \citep{Zhang2003} and some disk CMEs missed the earth
\citep[e.g.,][]{Schwenn2005,Wang2006a}; (3) Our model here gives
that MC deflection only happens during the process of shock front
penetrating MC body. The effect of shock pushing MC aside leads to
the deviation of MC by several degrees at the most; (4) We
conjecture that interaction between ICMEs may also be a cause of
ICME deflection, and the deflection angle could be up to tens
degrees, larger than that in (3). The propagation trajectory of
CMEs mentioned above is deflected from an initial straight line in
the IP medium. Both deflections in (1) and (2) are caused by
interaction between ambient solar wind and IP disturbance. In
contrary, the deflection in (3) and (4) are ascribed to
interaction between different IP disturbances, i.e. the collision
between MC-shock or MC-MC. It may expect a significant effect on
the possibility of CME hitting the earth in (1), (2), and (4),
whereas the effect in (3) may be negligible because of the small
deflection angle.

The deflection of shock aphelion in IP medium is a key factor in
the near-earth prediction of shock arrival time.
\citet{Hu1998,Hu2001} stated that the deflection of shock aphelion
results from joint effects of spiral IMF and heterogenous medium
consisting of fast and slow solar wind. The deflection is also
found here in OC of MC-shock. Starting from shock passage through
MC medium, shock aphelion deflects toward the contrary trend of MC
deflection until the shock totally merges with the MC-driven
shock. The final shock aphelion as well as front morphology are
distinct from those of isolated shock event. Both MC and shock
undergo significant modification during the process of their
collision.

\section{Concluding Remarks and Discussions}\label{Sec:Conclusion}
For further understanding of the IP ``shock overtaking MC" events
\citep{Wang2003b,Berdichevsky2005}, the investigation of MC-shock
interaction and consequent geoeffectiveness in paper 1 is
continued by a 2.5-dimensional ideal MHD numerical model. The
simulations find that shock eruption orientation relative to
preceding MC propagation plays a crucial role in MC-shock
interaction.

Firstly, MC-shock dynamical interaction is modeled. In order to
reveal the effect of the shock orientation relative to preceding
MC propagation, DC in paper 1 is here modified to be OC for
MC-shock interaction under the condition of the same shock speed.
The results show that the shock front in MC-shock OC behaves as a
smooth arc in MC medium. The cannibalized part of MC is highly
compressed by the shock along its normal. As the shock propagates
gradually into the preceding MC body, the most violent interaction
is transferred sideways (in terms of heliolatitude) with an
accompanying significant narrowing of the MC's angular width. The
opposite deflections of MC body and incidental shock aphelion
concur during the process of shock penetrating MC. MC deflection
ends when the shock approaches MC head; Shock deflection stops
when the shock completely merges with MC-driven shock. After shock
passage the MC is restored to oblate morphology. The high speed
flow right after MC inner boundary mentioned in paper 1 does not
exist here on the condition of non-uniform orientation of initial
MC and shock eruption.

Secondly, the geoeffectiveness of MC-shock OC is studied. Geoeffectiveness of an individual MC is largely
enhanced by an incidental ``non-central" shock. With the decrease of MC-shock commencement interval, shock
front at 1 AU traverses MC body and is responsible for the same change trend of the latitude of the
greatest geoeffectiveness of MC-shock compound. Among all cases with penetrating shock at various stages,
the maximum geoeffectiveness occurs when the shock enters MC core right at 1 AU. \citet{Wang2003c}
suggested that the maximum geomagnetic storm be caused by shock penetrating MC at a certain depth, and the
stronger the incident shock is, the deeper is the position. Based on our numerical model, Wang's conclusion
of shock penetration depth regarding the maximum geoeffectiveness \citep{Wang2003c} may be supplemented
that shock position is right at MC core on the condition of very strong shock.

Thirdly, the reliance of MC deflection on shock orientation and intensity is explored. The angular
displacements of MC body and shock aphelion are ascribed to MC-shock OC. An appropriate angular difference
between the initial eruption of an MC and an overtaking shock leads to the maximum deflection of the MC
body. The larger the shock intensity is, the greater is the deflection angle. The interaction of MCs with
other disturbances could be a cause of ICME's deflected propagation.

\begin{acknowledgments}
This work was supported by the National Natural Science Foundation
of China (40336052, 40404014, 40525014 and 40574063), and the
Chinese Academy of Sciences (startup fund). M. Xiong was also
supported by Innovative Fund of University of Science and
Technology of China for Graduate Students (KD2005030).
\end{acknowledgments}


%
%
%

\section*{Figure Captions}
\begin{description}
\item[Figure 1] The evolution of shock overtaking MC for Case
B$_2$, with (a)-(c) magnetic field magnitude $B$, (d)-(f) radial
flow speed $v_r$, and (g)-(i) radial characteristic speed of fast
mode $c_f$. Below each image are two additional radial profiles
along Lat.$=0^\circ$ and $4.5^\circ$S. Note: radial profile of $B$
is plotted by subtracting initial ambient value $B|_{t=0}$. The
white solid line in each image denotes the MC boundary. Solid and
dashed lines at each profile denote MC core and boundary.

\item[Figure 2] The evolution of shock overtaking MC for Case
C$_2$ with radial flow speed $v_r$. Only part of domain is
adaptively plotted to highlight MC.

\item[Figure 3] The time dependence of MC parameters: (a) radial
distance of MC core $r_m$, (b) MC radial span $Sr$ , (c) MC
angular span $S\theta$, (d) MC cross section area $A$, and (e) MC
core deflection angle $D\theta_m$. The solid, dashed and
dashed-dotted curves denote individual MC event (Case A), MC-shock
events (Case C$_2$, C$_1$). Three vertical delimiting lines
(dotted, dashed and dotted) from left to right correspond to the
occasion of shock encountering MC tail, core and head
respectively.

\item[Figure 4] The montage of radial flow speed $v_r$ for the
evolution of MC-shock compound at L1 under three conditions.

\item[Figure 5] The comparison of latitudinal distribution of
$Dst$ index among individual MC event (Case A) and MC-shock events
(Cases B$_2$ and C$_2$). The solid, dashed, and dashed-dotted
lines denote Case A, B$_2$, C$_2$ respectively, with the mark
$\Delta$, $\diamond$ for the passage of MC core and boundary. The
positive and negative latitude are referred to southern and
northern semi-heliosphere.

\item[Figure 6] The parameter variances of MC-related
geoeffectiveness as a function of $d_{Dst}$ in Group OC. Here
$d_{Dst}$ refers to radial distance between shock front and MC
inner boundary along sun-MC core. From left to right, three
vertical lines (dotted, dashed, dotted) denote the occasions of
shock just reaching MC tail, core, and front at L1 respectively.
The mark $\diamond$ and $\Delta$ denote corresponding
results of Case B$_2$ and C$_2$. (a) $Dt$, MC-shock emergence
interval, (b) $Dst$ index, (c) $~\mbox{Min.} (VB_z)$, the minimum
of dawn-dusk electric field $VB_z$, (d) $\Delta t$, the interval
between the commencement of $VB_z<-0.5~\mbox{ mV/m}$ and the
corresponding $Dst$ minimum, (e) $~\mbox{Min.}(Bs)$, the minimum
of southward magnetic component, and (f) $~\mbox{Max.}(B)$, the
maximum of magnetic magnitude. Solid, dashed and dashed-dotted
lines in (b) to (f) correspond to observations along
$~\mbox{Lat.=}0^\circ$, $4.5^\circ$S and $4.5^\circ$N
respectively.

\item[Figure 7] The response of the latitude of maximum
geoeffectiveness and accompanying $Dst$ (dashed line in (b)) as
the change of MC-shock interval $Dt$ in Group OC. The latitudinal
distribution of individual MC event (Case A) is denoted in solid
line of (b) as background. The relative ratio of geoeffectiveness
enhancement (c) by the shock is derived from two curves difference
of (b).

\item[Figure 8] The dependence of MC core deflection angle
$D\theta_m$ and $Dst$ at the specific latitude accompanying MC
core passage, on shock eruption orientation $\theta_{sc}$ (Group
SOD) and speed $v_s$ (Group SID). The horizontal dashed lines in
(b) and (d) denote corresponding $Dst$ of individual MC event
(Case A).

\end{description}

\begin{table}
\caption{Assortment of simulation cases of individual MC and
MC-shock interaction}\label{Tab1}
\begin{tabular}{|l|l|l|l|l|l|}
\hline
Group & Case & $v_s$(km/s)  &$\theta_{sc}(^\circ)$  & $t_{s0}$(hour) & Comment\\
\hline
IM & A & - & -& - & Individual MC \\
\hline  
DC & B$_1$, C$_1$, D$_1$, E$_1$, & 1630 & 0 & 41, 10, 60,
50, & Direct Collision \\
& F$_1$, G$_1$, H$_1$, I$_1$, &&& 48, 46, 44,
38, & \\
& J$_1$, K$_1$, L$_1$, M$_1$, &&& 35, 32,
29, 26, & \\
& N$_1$, O$_1$, P$_1$, Q$_1$ &&& 23, 20,
15, 6, & \\
& R$_1$ &&& 3 & \\
\hline  
OC & B$_2$, C$_2$, D$_2$, E$_2$, & 1630 & 10 & 41, 10, 60,
50, & Oblique Collision \\
& F$_2$, G$_2$, H$_2$, I$_2$, &&& 48, 46, 44,
38, & \\
& J$_2$, K$_2$, L$_2$, M$_2$, &&& 35, 32,
29, 26, & \\
& N$_2$, O$_2$, P$_2$, Q$_2$ &&& 23, 20,
15, 6, & \\
& R$_2$ &&& 3 & \\
\hline  
SOD & P$_1$, a, P$_2$, b, & 1630 & 0, 5, 10, 15, & 10 & Shock Orientation \\
& c, d, e, f, && 20, 25, 30, 40 &&Dependence \\
& g && 45 && \\
\hline  
SID & h, i, P$_2$, j, & 947, 1226, 1402, 1630, & 10 & 10 & Shock Intensity \\
& k, l, m, n &1773, 1997, 2314, 2686, & &&Dependence  \\
& o  & 3173 &&&\\
\hline  
\end{tabular}
\end{table}

\clearpage
\newpage

\newpage

\newpage

\newpage

\newpage

\newpage

\newpage

\newpage

\clearpage

\end{article}

\end{document}